\documentclass[lettersize,journal]{IEEEtran}
\pdfoutput=1
\usepackage{amsmath,amsfonts}
\usepackage{algorithmic}
\usepackage{algorithm}
\usepackage{array}
\usepackage[caption=false,font=normalsize,labelfont=sf,textfont=sf]{subfig}
\usepackage{textcomp}
\usepackage{stfloats}
\usepackage{url}
\usepackage{verbatim}
\usepackage{graphicx}
\usepackage{cite}
\usepackage{multirow}
\usepackage{hyperref}
\usepackage{booktabs} 
\usepackage{caption}

\hypersetup
{hypertex=true,
colorlinks=true,
linkcolor=black,
anchorcolor=black,
citecolor=black
}

\begin{document}

\title{Light Field Imaging in the Restrictive Object Space\\ based on Flexible Angular Plane}

\author{Ping Zhou, Nuo Chen, Yuda Xu, and Chengcai Xu

    \thanks{Ping Zhou, Nuo Chen, Yuda Xu, and Chengcai 
    Xu are with the School of Biological Science and Medical Engineering, Southeast University, Nanjing 210096, China (e-mail:capzhou@163.com)}
}


\IEEEpubid{\begin{minipage}{\textwidth}\ \\[30pt] \centering
    0000--0000/00\$00.00~\copyright~2021 IEEE
\end{minipage}}
\IEEEpubidadjcol

\maketitle

\begin{abstract}
In some applications, the object space of light field imaging system is restrictive, such as industrial and medical endoscopes. If the traditional light field imaging system is used in the restrictive object space (ROS) directly but without any specific considerations, the ROS will lead to severe microlens image distortions and then affects light field decoding, calibration and 3D reconstruction. The light field imaging in restrictive object space (ROS-LF) is complicated but significant. 
In this paper, we first deduce that the reason of the microlens image deviation is the position variation of the angular plane, then we propose the flexible angular plane for ROS-LF, while in the traditional light field the angular plane always coincides with the main lens plane.
Subsequently, we propose the microlens image non-distortion principle for ROS-LF
and introduce the ROS-LF imaging principle. We demonstrate that the difference is an aperture constant term between the ROS-LF and traditional light field imaging models. At last, we design a ROS-LF simulated system and calibrate it to verify principles proposed in this paper.

\end{abstract}


\section{Introduction}
\IEEEPARstart{A}{s} the collection of light rays emitted from object space, light fields record both the intensity of light rays and their angular information. Recently, light field image processing has drawn extensive research interest due to its tremendous development, and has been widely used in industrial, medical applications and so on \cite{light_field_cameras,light_field_microendoscopy,Plenoptic_cameras,Convolutional_Networks}.

Light field imaging principles are fundamental to decoding, 3D reconstruction algorithms and applications. Generally, light fields are represented by the two-parallel plane (TPP) model\cite{levoy_light_1996}, where a light ray is described by intersections with two planes. In the typical microlens array (MLA) based light field imaging system, a microlens is often interpreted as an image pixel in the traditional imaging system, and a photosensor pixel in microlens image is considered as one of the many light rays that contribute to the image pixel\cite{Ng's_thesis}. In previous researches, the main lens plane is generally defined as the angular plane and the MLA plane as the spatial plane\cite{levoy_light_1996,calibration_method1,calibration_method2,calibration_method3}. Many works based on the TPP model have been constructed. For example, Dansereau et al. presented a 15-parameter plenoptic camera model to decode pixels into rays, and a 12-free-parameter transformation matrix for calibration \cite{dansereau_decoding_2013}, Q. Zhang et al. presented a generic multi-projection-center (MPC) model with 6 intrinsic parameters to characterize light field cameras with different image formations \cite{zhang_generic_2018}.

\par
\begin{figure}[t]
\centering

\includegraphics[width=3in]{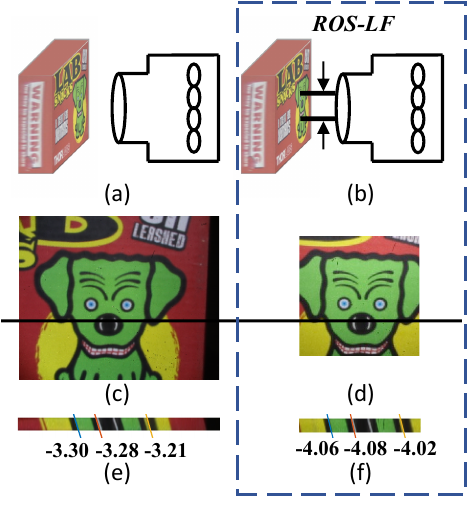}
\captionsetup{labelsep=period, font=small} 
\caption{The line slope difference in the traditional LF and ROS-LF. (a) Imaging in the traditional LF. (b) Imaging in the ROS-LF. (c) Central sub-aperture image by the traditional LF. (d) Central sub-aperture image by the ROS-LF. (e) Some line slopes in EPI for (c). (f) Some line slopes in EPI for (d).}
\label{figDiffEPI}
\end{figure}

Although many works on light field model, decoding, calibration and 3D reconstruction have been presented, they all focused on light rays in the free object space of the main lens. In other words, the main lens of light field imaging system receives light rays in its object space from all directions. However, in some applications such as industrial and medical endoscopes, the object space of the main lens is somehow restricted. In the light field in restrictive object space (ROS-LF), the main lens receives light rays only from some finite directions. 
The restricted object space results in severe microlens image distortions visually\cite{schambach_microlens_2020}, including deviations from the true perspective projected centers and deformations (characteristic cat eye shape microlens images), in particular in off-center microlens images. For distorted microlens images, they are not only challenging to detect accurate microlens image centers\cite{schambach_microlens_2020}, but also lead to unexpected error in light field decoding and applications. As shown in Figure \ref{figDiffEPI}, we capture two light field images with a traditional light field camera in free and restrictive object space respectively, where the distance between the camera and the object are fixed. It seems that the only difference lies on the imaging regions as parts of light rays cannot reach the light field camera due to the ROS, as shown in Figure \ref{figDiffEPI} (c) and (d). However, the most significant issue is that in the EPIs, the line slopes for the same points are different indeed, as shown in Figure \ref{figDiffEPI} (e) and (f), where line slopes of three points are shown. Therefore, if the traditional light field camera is used in ROS directly but without specific considerations, we will obtain wrong depth estimation results as the line slope is related to the depth \cite{Zhou_EPI,2017_EPI,shared_EPI,Liu:17,EPI_Patch}.

Unfortunately, most literature focused on light field imaging model, decoding methods and applications, but paid little attention to the reason of the microlens image distortion and how to solve it. In this paper, the main purpose for ROS-LF imaging is to explore ROS-LF imaging principles and design ROS-LF system with non-distorted microlens images, rather than estimate microlens image centers from distorted microlens images\cite{schambach_microlens_2020}. In particular, our contributions are as follow:

1) Based on TPP principles, we explore that the reason of microlens image deviations is the variation of the angular plane. Then, the flexible angular plane (FAP) is proposed for ROS-LF, which is composed of a virtual aperture plane and an angular plane separated from traditional light field model. More details are depicted in Section II-A and B.

2) Based on the FAP, we propose the microlens image non-distortion and $f$-number matching principle for ROS-LF, which is used to guide the design of ROS-LF imaging system. More details are depicted in Section II-C and D.

3) In Section II-E, we propose the ROS-LF model and find the difference lies on an aperture constant term between the ROS-LF and traditional LF models, which is demonstrated by a simulated ROS-LF imaging system. Furthermore, a calibration method for ROS-LF system is proposed.

\section{ Methods}
Before presenting the ROS-LF analysis, we first make a brief introduction about the terms that are defined and commonly used in this paper, as listed in Table \ref{table1}. 

\begin{table}[h]
    \centering
    \caption{Symbols in ROS-LF imaging model}
    \begin{tabular}{c|c}
        \hline
\textbf{Term}&\textbf{Definition}\\ \hline
$(O_F,X_F,Y_F,Z_F)$& FAP coordinates system\\ \hline
$(O_m,X_m,Y_m)$& MLA coordinate system\\ \hline
$(O_p,x,y)$ & Photosensor coordinate system \\ \hline
$L(s,t,u,v)$ & Decoded light field in the unit of pixel\\ \hline
    \end{tabular}
    \label{table1}
\end{table}

Without loss of generality, the optical center and optical axis of the FAP plane are defined as the origin $O_F$ and the $Z_F$-axis of the FAP coordinate system. Based on the TPP principle, the FAP, MLA plane and the photosensor plane are all perpendicular to the $Z_F$-axis.

\subsection{Analysis of Microlens Image Deviations}
Based on TPP principles of light field, a microlens image should be thought as an image pixel in the spatial plane, and a photosensor pixel value in the microlens image should be thought as one of the many light rays that all contribute to that image pixel but pass through the angular plane with different angles. In other words, microlens images record the structure of light in the scene, represent not only the spatial information but also the angular information.

\begin{figure}[htbp]
\centering
\includegraphics[width=3.3in]{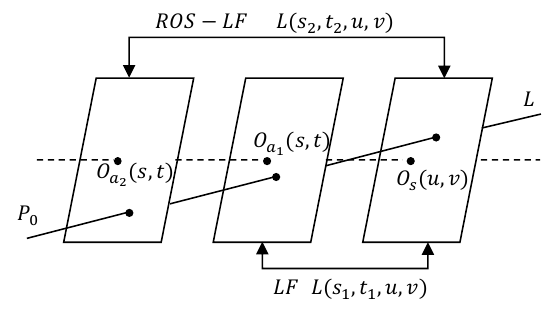}
\captionsetup{justification=centering}
\captionsetup{labelsep=period, font=small} 
\caption{The ROS-LF model based on TPP}
\label{figtwoTPP}
\end{figure}

For a specific point $P_0$ in object space of an optical imaging system, as shown in Figure \ref{figtwoTPP}, a light ray that contributes to $P_0$ terminates the identical microlens on the MLA plane based on optical imaging principles, whether $P_0$ is in a traditional light field or a ROS-LF. While in the ROS-LF, there exists microlens image deviations with respect to the traditional light field.
What does microlens image deviation mean in ROS-LF? On one hand, since the spatial information remains constant whether $P_0$ is in a traditional light field or a ROS-LF, it means that the spatial plane in TPP principles remains constant. On the other hand, as microlens images record angular information of the ROS-LF or traditional LF, it is reasonable to assume that the angular plane of ROS-LF varies as the microlens image deviation occurs, as shown in Figure \ref{figtwoTPP}. In other words, compared to the traditional light field, microlens image deviation in ROS-LF corresponds to changing the position of the angular plane, and leads to the line slope changing in EPI, as shown in Figure \ref{figDiffEPI} (e) and (f).
\par

As shown in Figure \ref{figtwoTPP}, $O_s (u,v)$ is a spatial plane, $O_{a_1}(s,t)$ and $O_{a_2}(s,t)$ are two parallel angular planes and are both parallel to the spatial plane. Let $L(s,t,u,v)$ denote the light travelling along the ray that intersects angular plane at $(s,t)$ and spatial plane at $(u,v)$. As shown in Figure \ref{figtwoTPP}
, the light ray $L$ emitter from $P_0$ intersects the spatial plane $Os (u,v)$ at $(u,v)$, the angular planes $O_{a_1}(s,t)$ and $O_{a_2}(s,t)$ at $(s_1,t_1 )$ and $(s_2,t_2)$, respectively. If $O_{a_1}(s,t)$ and $O_s (u,v)$ construct a traditional light field, while $O_{a_2}(s,t)$ and $O_s (u,v)$ construct a ROS-LF. Figure \ref{figtwoTPP} illustrates how a light ray $L$ parameterized by the angular plane $O_{a_1}(s,t)$ for $L(s_1,t_1,u,v)$ may be re-parameterized by its another intersection with the angular plane $O_{a_2}(s,t)$ for $L(s_2,t_2,u,v)$. As shown in Figure \ref{figtwoTPP}, changing the position of the angular plane corresponds to changing the angular description of the 4D light field only. In these two descriptions of different 4D light field, the light ray $L$ contributes to the same spatial coordinates, but varies in the angular coordinates, which coincides with the difference between ROS-LF and traditional LF. Therefore, when the traditional LF is used for ROS imaging, the reason of microlens image deviation is that the angular plane position changes. Then, how does the angular plane of ROS-LF change?

\subsection{Angular Plane Analysis in ROS-LF}
Let us focus on the definition of angular plane in light field based on TPP principles.
\par
\begin{figure*}[!t]
\centering
\subfloat[\label{fig2(a)}]{\includegraphics[width=3.5in]{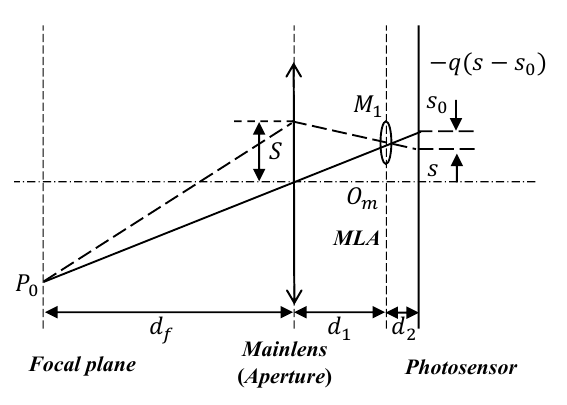}}%
\hfil
\subfloat[\label{fig2(b)}]{\includegraphics[width=3.5in]{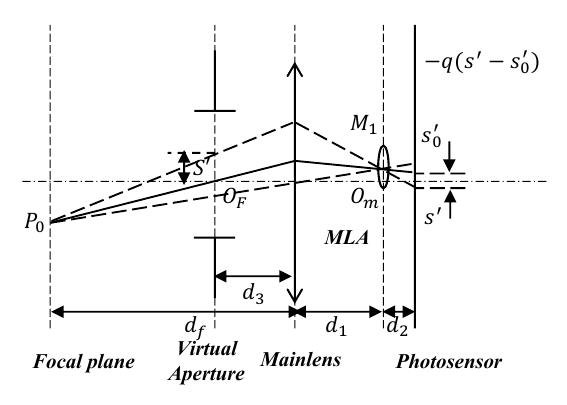}}%
\captionsetup{labelsep=period, font=small} 
\captionsetup{justification=centering}
\caption{The analysis of angular plane in light field. (a) Traditional LF. (b) ROS-LF.}
\label{figAnalysis_AP}
\end{figure*}

In traditional LF, we usually define the main lens plane as the angular plane \cite{levoy_light_1996}. However, there exists an aperture plane that is integrated in the main lens but is neglected by many works. Actually, we define the main lens plane and the aperture plane together as the angular plane of LF \cite{Ng's_thesis}, as shown in Figure \ref{figAnalysis_AP}(a).
It is worth to note that the microlens scale in this figure is somewhat deceptive, because we adjust it artificially to make it possible to see them clearly.  In the object space of the main lens, for a point $P_0$ on its focal plane, light rays emitted from $P_0$ pass through the main lens and converge at the microlens $M_1$. The center of its corresponding microlens image is $s_0$, that is determined by the light ray passing through the center of the angular plane, as shown by the black bold line in Figure \ref{figAnalysis_AP}(a). It’s worth to note that the centers of the angular plane, the main lens plane and the aperture plane are coincided in traditional LF.
\par
Based on light field principles, microlens images record angular information of light field. For a point $P_0$, the angular information illustrates the relationship between a pixel with coordinate of $s$ in its corresponding microlens image and the position of $S$ where a light ray emitted from $P_0$ intersects the angular plane, as shown in Figure \ref{figAnalysis_AP}(a). Therefore, there exists 
\begin{equation}
S=-\frac{qd_1}{d_2}(s-s_0)\label{Q1}
\end{equation}

\par

where $d_1$ is the distance between main lens and MLA, $d_2$ is the distance between MLA and photosensor. As shown in Figure \ref{figAnalysis_AP}(a), the object space of the main lens is free in the traditional LF, all light rays emitted from $P_0$ could reach the main lens. However, in ROS-LF, 
as some light rays cannot reach the main lens, there seems exist a virtual aperture in its object space, which obstructs these light rays. To facilitate the ROS-LF imaging analysis, we refer to the distributed and lumped parameter analysis, and consider the ROS as a virtual aperture in the ROS-LF, whose entrance pupil and position are variable according to details of the ROS, as shown in Figure \ref{figAnalysis_AP}(b). Furthermore, it is unsuitable to define angular plane with the main lens and the integrated aperture together now. Then, how to define the angular plane in ROS-LF? 
\par
As shown in Figure \ref{figAnalysis_AP}(b), light rays emitted from $P_0$ are symmetric to the virtual aperture’s optical axis, but asymmetric to the main lens’s optical axis as parts of light rays are occluded by the virtual aperture. Therefore, it is the aperture but not the main lens that is essential to determine microlens images. To analysis ROF-LF and obtain circular (symmetric) microlens images, we define the virtual aperture as the flexible angular plane (FAP) of ROS-LF, where ‘Flexible’ means the size and position of the virtual aperture are variable according to the ROS details. In ROS-LF, the main lens is just the optical imaging component to focus on $P_0$ at a desired microlens. As shown in Figure \ref{figAnalysis_AP}(b), we parameterize the FAP by $(s',t')$, so that $L(s',t',u,v)$ is the light ray travelling through the FAP and the traditional spatial plane. With the FAP definition, it is well known based on TPP principles that the angular coordinate $S'$ appearing on the FAP but not the traditional angular plane is given by 
\begin{equation}
S'=-\frac{1}{H}\frac{qd_1}{d_2} (s'-{s_0}') \label{Q2}
\end{equation}
and
\begin{equation}
 H=\frac{d_f}{d_f-d_3}\label{Q3}
\end{equation}

where ${s_0}'$ is the center of microlens image in ROS-LF, which is determined by the light ray passing through the center of the FAP, as shown by the black bold line in Figure \ref{figAnalysis_AP}(b). $d_f$ is the distance between the main lens plane and its focal plane, $d_3$ is the distance between the main lens plane and FAP. According to the FAP definition, $d_3$ is the position of FAP and is related to ROS details. It is obvious that, in form at least, the angular relationship in ROS-LF is similar to that in the traditional LF, but there exists a constant difference of $H$ between them. In this paper, we define it as the \textbf{Aperture Constant} that is related to $d_3$ and $d_f$, as depicted in Eq.(\ref{Q3}). According to the FAP definition, Aperture Constant is related to ROS details also.
\par
To demonstrate our assumption about FAP in ROS-LF, we deduce the center deviation $\Delta x$ of microlens image between the ROS-LF and the traditional LF based on light field principles, that is expressed as follow  
\begin{equation}
 \Delta x=\frac{d \cdot d_2 \cdot  d_3 \cdot d_f }{q \cdot {d_1}^2 \cdot (d_f-d_3)}~(x-x_0)\label{Q4}
\end{equation}

where $d$ is the pitch of microlens, $x_0$ is the coordinate of the intersection of the photosensor plane and the optical axis in 2D photosensor plane $(x, y)$. According to Eq.(\ref{Q4}), the farther the microlens image is from the center of the photosensor plane, the more severe the microlens image deviation is, which is consistent with the experimental results of Schambach M. \cite{schambach_microlens_2020}.

\begin{figure}[t]
\centering
\includegraphics[width=3in]{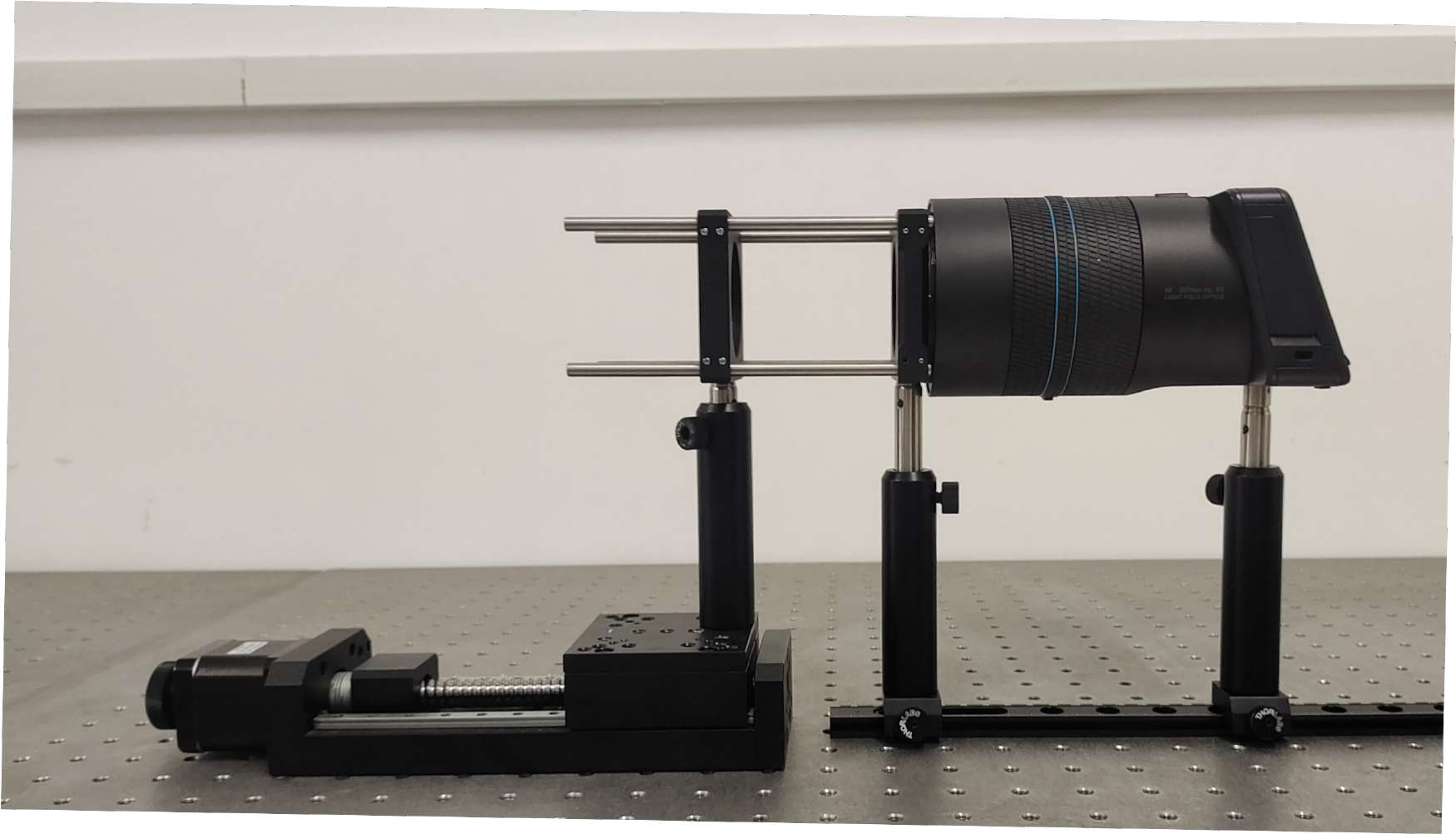}
\captionsetup{labelsep=period, font=small} 
\captionsetup{justification=centering}
\caption{The Simulated ROS-LF Imaging System}
\label{figSimuROSLF}
\end{figure}

\begin{figure}[t]
\centering
\includegraphics[width=3in]{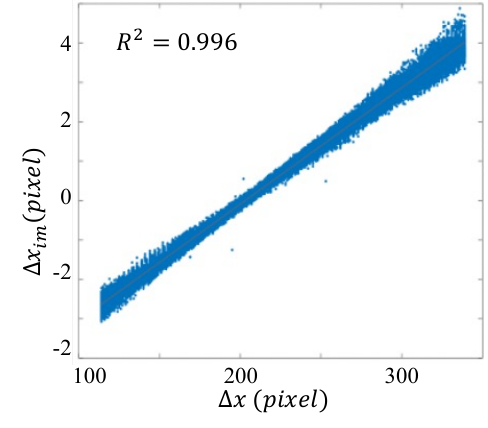}
\captionsetup{labelsep=period, font=small} 
\caption{Verification of the FAP assumption. }
\label{figFAPVerify}
\end{figure}

Before the FAP assumption is verified, the imaging system cannot be calibrated based on it. Although some parameters in Eq.(\ref{Q4}) are unknown, it’s certain that $\Delta x$ is linear to $x$. 
A simulated ROS-LF imaging system is designed to verify the FAP assumption, as shown in Figure\ref{figSimuROSLF}. The system is equipped with a traditional light field camera (Illumn, Lytro II) and an aperture produced by 3D printer that is used as the virtual aperture to simulate ROS-LF. The light field camera and aperture are mounted in kinematic mounts allowing fine adjustment in respect to the optical axis. The aperture is placed in the object space of the main lens, and a translation stage is used to adjust the distance $d_3$ between the aperture and main lens. The relationship between $\Delta x$ and $x$ is depicted as Figure\ref{figFAPVerify}, and the linear fitting error is 0.996. It is worth noting that there are more than one microlens image for one $x$-axis coordinate, as shown in Figure \ref{figFAPVerify}. The experimental results demonstrate that the “flexible angular plane” hypothesis proposed in this paper is right.

\subsection{Microlens Image Non-distortion Principle}

In the sub-sections above, we have analyzed the reason of microlens lens deviations and proposed the FAP principle in ROS-LF. In this sub-section, we will analyze the microlens image deformation and introduce the microlens image non-distortion principle based on the FAP. Obviously, microlens image is regular when object space is free in the traditional light field. While in the ROS-LF, before light rays emitted from a certain point $P_0 (X_0,Z_0)$ reach the photosensor, occlusions may occur due to the restrictive object space and the aperture of the main lens, which lead to unexpected microlens image distortions. Although ROS are inevitable in the ROS-LF, as long as light rays passing through virtual aperture and main lens are symmetric in respect to the center $O_F$ of the FAP, the microlens image is still regular and non-distorted.

\begin{figure}[t]
\centering
\includegraphics[width=3.3in]{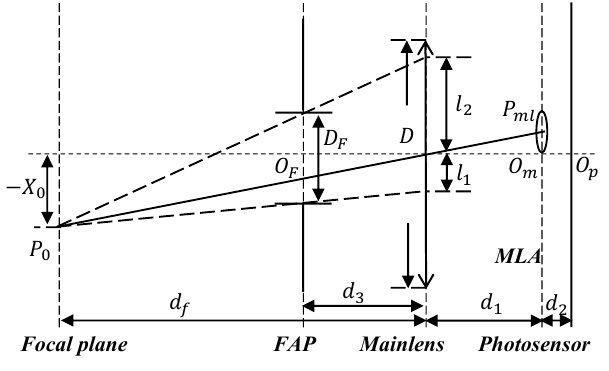}
\captionsetup{labelsep=period, font=small} 
\captionsetup{justification=centering}
\caption{Microlens Image Non-distortion Principle }
\label{figNondistortion}
\end{figure}

As shown in Figure \ref{figNondistortion}, for light rays passing through the whole virtual aperture\cite{Yang_distortionless}, there is relationship based on light field principles, which is expressed as follows
\begin{equation}
\frac{D_F}{2}=-\frac{d_3}{d_f} X_0+\frac{(d_f-d_3)}{d_f}l_2\label{Q5}
\end{equation}
and\
    $$\frac{X_0}{X_{ml}}=\frac{d_f}{d_1}$$

where $D_F$ is the pupil size of the virtual aperture, $X_0$ is the coordinate in $x$-axis of $P_0$, $l_1$  and $l_2$ are the $x$-axis coordinates of the intersection between the main lens and the rays passing through the upper and lower edges of the aperture stop. When light ray is not occluded by the main lens, that is $l_2\le D⁄2$, all light rays emitted from $P_0$  could reach the MLA without any occlusion by the main lens, substituting it into Eq.(\ref{Q5}), there is
\begin{equation}
\frac{X_0}{d_f}\le\frac12{\left(\frac{D-D_F}{d_3}-\frac D{d_f}\right)}.\label{Q6}
\end{equation}

where $D$ is the diameter of the main lens aperture. This is the microlens image non-distortion principle in ROS-LF. As depicted in Eq.(\ref{Q6}), $D_F$ and $d_3$ are related to the virtual aperture (i.e., the ROS), $D$ and $d_f$ are related to the main lens of the traditional light field imaging system, $X_0$ is the coordinate of measured point $P_0$ in $x$-axis. Therefore, the microlens image non-distortion principle is related to the ROS circumstance, the light field imaging system parameters and the position of the measured object. On one hand, inappropriate parameters result in severe distorted microlens images, in particular in off-center microlens images. On the other hand, Eq.(\ref{Q6}) illustrates that there is also a field of view (FOV) determined by both the ROS circumstance and imaging system parameters. In FOV, objects are imaged with non-distorted microlens images. 

\subsection{$f$-number Matching}
To cover as many pixels on photosensor as possible, the $f$-number of the main lens has to be equal to that of the microlens in traditional light field camera. The $f$-number matching should be met in ROS-LF imaging system also. In traditional light field camera, the $f$-number of main lens is the ratio of the intrinsic focal length to the entrance pupil diameter. In ROS-LF, the $f$-number of the microlens is still identical to that in traditional light field, but the $f$-number of “main lens” is changed due to the virtual aperture (i.e., ROS).
\begin{figure}[t]
\centering
\captionsetup{labelsep=period, font=small} 
\includegraphics[width=3.3in]{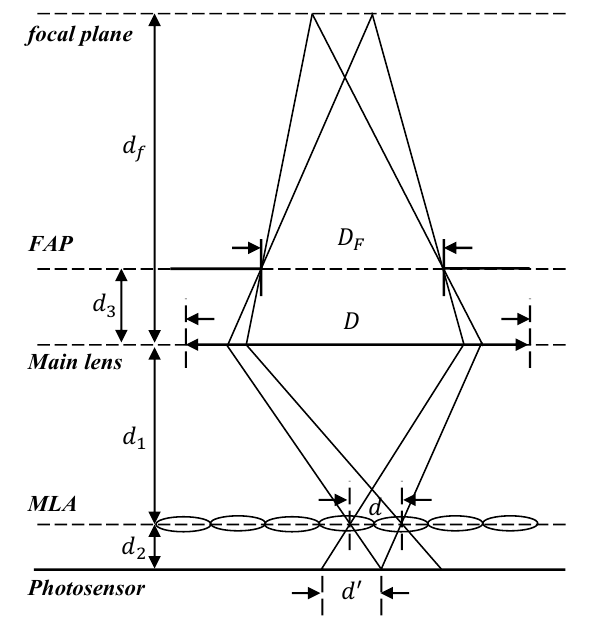}
\captionsetup{justification=centering}
\caption{$f$-number matching in ROS-LF }
\label{figFnumber}
\end{figure}

As shown in Figure \ref{figFnumber}, there are two adjacent microlens images and their corresponding points on the focal plane of main lens. When these two adjacent microlens images are tangent, the $f$-number matching principle is derived based on the similar triangle principles, which is expressed as follow
\begin{equation}
\frac{d_1\cdot D_{F}+d_3\cdot d}{(d_f-d_3)\cdot d\cdot d_1}=\frac{d_1+d_2}{d_2\cdot d_f}
\label{Q7}
\end{equation}

In general, $d_1\cdot D_F$ is much larger than $d_3\cdot d$ ($d_1\cdot D_{F} \gg d_3\cdot d$), and $d_1$ is much larger than $d_2$ ($d_1 \gg d_2$), then Eq.(\ref{Q7}) is simplified as follow.
\begin{equation}
\frac{d_2}d\approx\frac{d_1\cdot(d_f-d_3)}{D_F\cdot d_f}=\frac{d_1}{D_F\cdot H}
\label{Q8}
\end{equation}

Compared to the $f$-number matching principle in the traditional light field as depicted in Eq.(\ref{Q9}), the $f$-number matching principle in ROS-LF is related to the ROS circumstance undoubtedly, as the diameter $D$ of the main lens is substituted with the product of the diameter $D_F$ of the virtual aperture and the aperture constant $H$. Furthermore, although Eq.\ref{Q8} is more complicated, it is similar with that of the traditional LF in form.

\begin{equation}
\frac{d_2}d=\frac{d_1}D
\label{Q9}
\end{equation}

\subsection{ROS-LF Imaging Model and Calibration}
Based on flexible angular plane principles, the ROS-LF imaging model is proposed in this paper, which is similar to that of traditional light field.
\begin{figure}[t]
\centering
\includegraphics[width=3.5in]{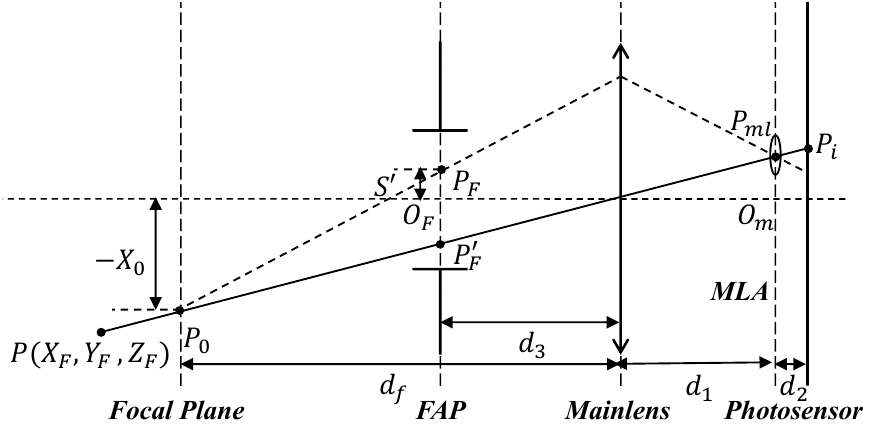}
\captionsetup{labelsep=period, font=small} 
\captionsetup{justification=centering}
\caption{ROS-LF imaging model }
\label{figROSmodel}
\end{figure}

As shown in Fig.\ref{figROSmodel}, consistent with the traditional light field, the ROS-LF imaging model also consists of a main lens, a MLA and a photosensor, whose optical axes coincide. The main lens is modeled as thin lens and the microlens is model as pinhole. In ROS-LF, it is different to the traditional light field that the angular plane is not the main lens plane, but the FAP defined above. To express the ROS-LF imaging procedure in mathematical form, the FAP is modeled as thin lens, and every microlens in the MLA is model as pinhole. Without loss of generality, a light ray emitted by $P$($X_F$,$Y_F$,$Z_F$) passes through the focal plane at $P_0$ and the main lens at $P_m$, heads to the point $P_{ml}$ of the MLA plane, and images at the point $P_i (X_i,Y_i)$ of the photosensor plane. The light ray and the FAP intersect at $P_F$ that is $S'$ from the center.

As points $P$, $P_0$, and $P_m$ are collinear, $\Delta P_0 P_F P_F'$ and $\Delta P_0 P_m O_m$, $\Delta O_{ml} P_{ml} O_m$ and $\Delta O_F P_F' O_m$ are both similar triangles, the ROS-LF imaging model is described as (\ref{Q10})
\begin{equation}
\begin{bmatrix}u\\v\\1\end{bmatrix}=\frac1{Z_F}\begin{bmatrix}-\frac{d_1}{d\cdot H}&0&x_0-\frac{S^{\prime}\cdot d_1}{d\cdot d_f}&\frac{S^{\prime}\cdot d_1}{d\cdot H}\\0&-\frac{d_1}{d\cdot H}&y_0-\frac{T^{\prime}\cdot d_1}{d\cdot d_f}&\frac{T^{\prime}\cdot d_1}{d\cdot H}\\0&0&1&0\end{bmatrix}\begin{bmatrix}X_F\\Y_F\\Z_F\\1\end{bmatrix}.
\label{Q10}
\end{equation}

where ($u,v$) is the coordinates of $P_{ml}$ in pixels. For comparison, the traditional LF imaging model is expressed as Eq. (\ref{Q11})
\begin{equation}
\begin{bmatrix}u\\v\\1\end{bmatrix}=\frac1{Z_{m}}\begin{bmatrix}-\frac{d_1}d&0&x_0-\frac{S\cdot d_1}{d\cdot d_f}&\frac{S\cdot d_1}d\\0&-\frac{d_1}d&y_0-\frac{T\cdot d_1}{d\cdot d_f}&\frac{T\cdot d_1}d\\0&0&1&0\end{bmatrix}\begin{bmatrix}X_{m}\\Y_{m}\\Z_{m}\\1\end{bmatrix}
\label{Q11}
\end{equation}

It is obvious that these two imaging models are similar in form, and the only difference lies on the aperture constant $H$ in some parameters in ROS-LF imaging model. 

To compute the parameters in ROS-LF imaging model, we propose an advanced two-step calibration method based on our previous works \cite{two_step_ZHOU2019190}. In the first step calibration, the parameters $d_1⁄H$, $(x_0,y_0)$ are derived through the central sub-aperture image by fixing $S'=T'=0$, where Zhang's calibration method is used \cite{Zhang_calibration}. The diameter $d$ of the MLA can be obtained from its metadata. Due to the change of the angular plane, the second step calibration of the ROS-LF is different from the traditional light field, but there is similar relationship expressed as follow:
\begin{equation}
\Delta x_{ml}=\frac qd\frac{d_1}H\frac1{d_2}\left(\frac{d_1}{d_f}\Delta x_{im}-\frac{d_1}H\frac{\Delta x_{im}}{Z_F}\right)
\label{Q12}
\end{equation}

For two points in the scene, $\Delta x_{ml}$ is the distance between their image points in a sub-aperture image, $\Delta x_{im}$ is the distance between their image points in a microlens image. $Z_F$ is the $Z$-axis coordinate of one of them and is obtained after the first step calibration. According to Eq. (\ref{Q12}), the parameters $d_2$, $d_1⁄d_f$  are calibrated by the least square fitting method. Furthermore, the parameters $d_1$ and $d_f$ are obtained according to the optical Gaussian formula. Finally, the parameters $H$ and $d_3$ are derived according to $d_1$ and $d_f$.


\section{Experimental Results}
\begin{figure}[t]
\centering
\includegraphics[width=3in]{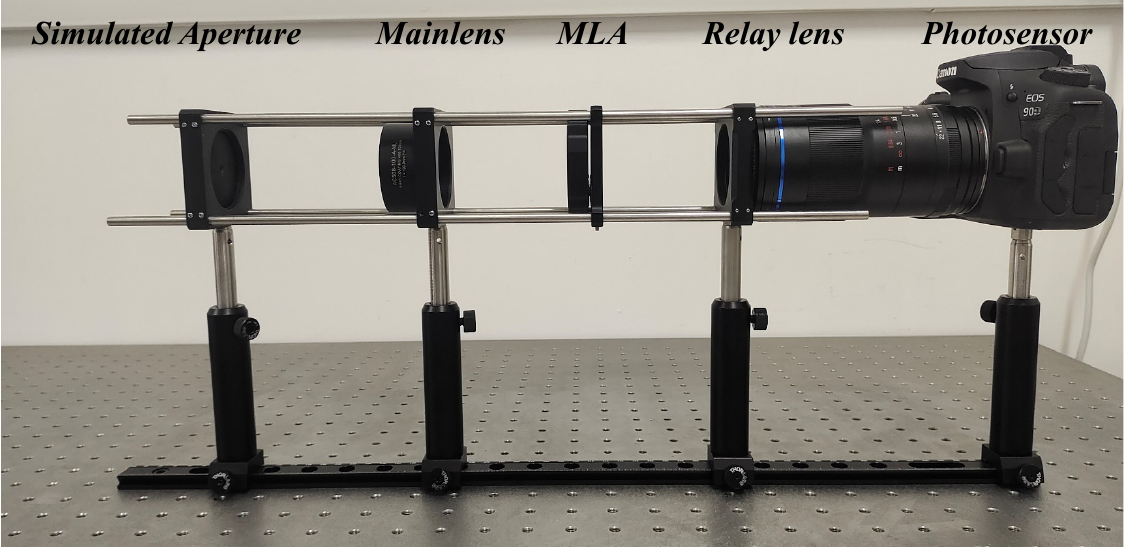}   
\captionsetup{labelsep=period, font=small} 
\captionsetup{justification=centering}
\caption{setup of the simulated ROS-LF system}
\label{figSetup_ROS}
\end{figure}

A simulated ROS-LF imaging system is shown in Figure \ref{figSetup_ROS}, which is equipped with a mainlens(Thorlabs, N-BK7, focal length $100mm$, diameter $50.8mm$), a MLA(AUS, pitch $63um$, focal length $2mm$), a 1:1 relay macro lens(LAOWA, focal length $100mm$) and a photosensor(Canon 90D, pixel size $3.2um$). Furthermore, a simulated aperture produced by 3D printer is placed in the object space of the mainlens, which is used to simulate the ROS. All components of the simulated ROS-LF system are mounted in kinematic mounts allowing fine adjustment in respect to the optical axis. The pupil size $D_F$ of the simulated aperture is $2.9mm$. The central sub-aperture image and microlens images are extracted via the Matlab LFToolbox (V0.4) proposed by Dansereau \cite{dansereau_decoding_2013,dansereau2015linear}. The size of raw LF image is $4600\times3080$ pixels, the size of the microlens image is $13\times13$ pixels so that the size of central sub-aperture image is $356\times237$ pixels. 
\par
\begin{figure}[t]
\centering
\subfloat[\label{fig9(a)}]{
\includegraphics[width=1.35in]{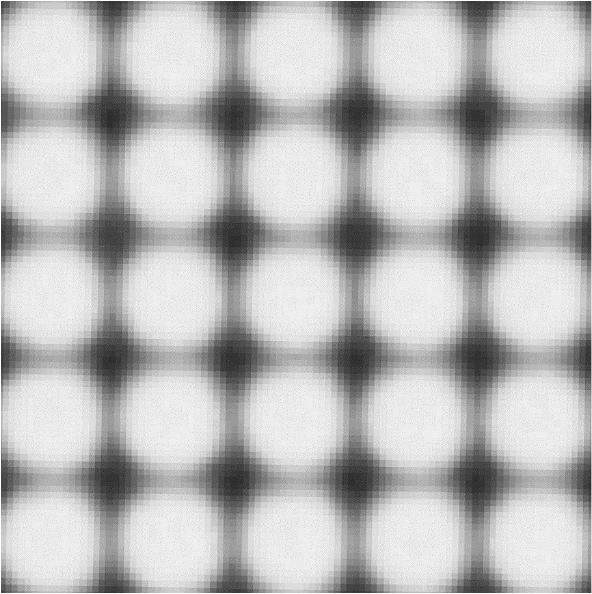}}
\hfil
\subfloat[\label{fig9(b)}]{
\includegraphics[width=1.35in]{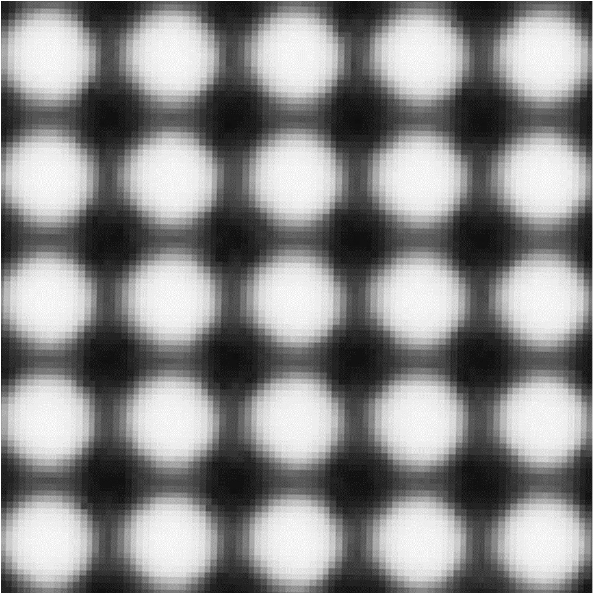}}

\subfloat[\label{fig9(c)}]{
\includegraphics[width=1.35in]{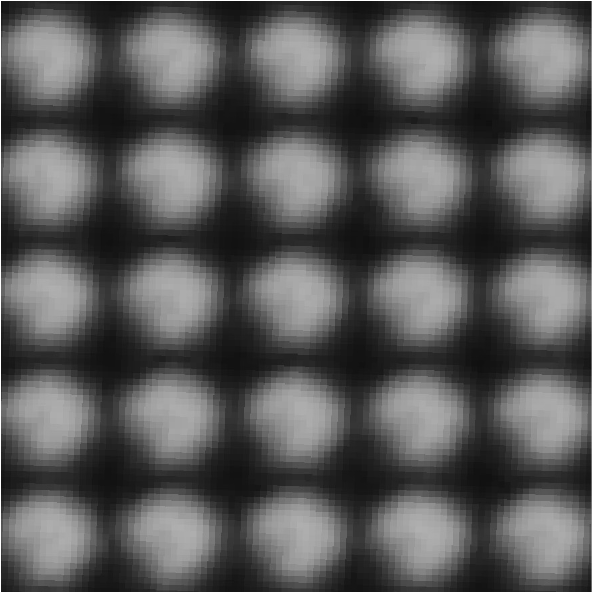}}
\hfil
\subfloat[\label{fig9(d)}]{
\includegraphics[width=1.35in]{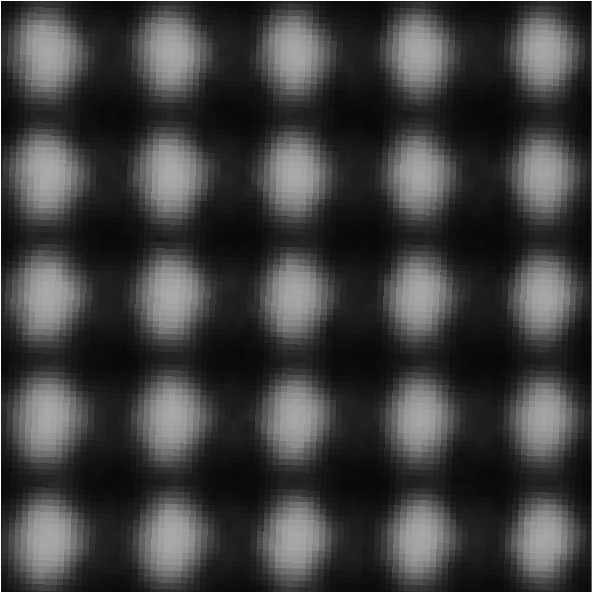}}
\captionsetup{labelsep=period, font=small} 
\captionsetup{justification=centering}
\caption{Microlens images under different parameters}
\label{figMI_differentPara}
\end{figure}

To verify the $f$-number matching and microlens image non-distortion principles, we adjust the parameters of the simulated ROS-LF system approximately as follows: $d_f \approx 350mm$, $d_1 \approx 140mm$, $d_2 \approx 2mm$ and $d_3 \approx 120mm$. Therefore, the left-side and right-side values of Eq.\ref{Q8} are approximately $31.75$ and $31.72$, respectively. It is obvious that the $f$-numbers of the MLA and FAP are matched, and some microlens images are shown in Figure \ref{figMI_differentPara}(a). On the other hand, the right-side of the Eq.\ref{Q9} is equal to 2.76 approximately, so that the $f$-numbers of the MLA and the mainlens are not matched as the traditional light field imaging system does. Futhermore, we adjust $d_3$ to $20mm$ by shifting the simulate aperture along the rail in Figure \ref{figSetup_ROS}, then the right-side value of Eq.\ref{Q8} is equal to 45.52 and the $f$-numbers are not matched, some microlens image are shown in Figure \ref{figMI_differentPara}(b), which are at the same position on the raw LF image with those in Figure \ref{figMI_differentPara}(a). The experimental results demonstrate that the $f$-number matching principle in ROS-LF is right. 
\par
For the microlens image non-distortion principle, according to Eq.\ref{Q6} and the approximate parameters of the simulated ROS-LF system ($d_3 \approx 120mm$), we obtain that when $X_0$ is less than $44.5mm$, the microlens images are not distorted. However, according to the optical principles, the FOV on the focal plane of the tradional light field system is approximately $57.5mm$ as the MLA is a $23mm$ square. When $X_0$ exceeds the non-distortion range, the microlens images closer to the MLA edge are distorted, as shown in Figure \ref{figMI_differentPara}(c) and (d). The experimental results also demonstrate that the microlens image non-distortion principle in ROS-LF is right. 
\par
We captured 12 checkerboard images to accomplish the simulated ROS-LF system calibration. According to the range of $X_0$, the checkerboard is a pattern of $4\times5$ cells, whose grid size is $6mm\times6mm$. In our experiments, the checkerboard is about $350mm$ away from the main lens. The advanced two-step calibration method proposed in Section II-E is implemented, and the nonlinear optimization is alos implemented by the Levenberg-Marquardt algorithm. The parameters copied from the metadata provided by production company and computed by our advanced two-step calibration method as shown in Table \ref{table2}.

\par
\begin{table}[t]
\centering
\caption{Calibration Results of the ROS-LF imaging system}
\begin{tabular}{ccc}
\toprule
   &Parameters&Values\\
\midrule
\multirow{5}*{The first step calibration}&$(x_0,y_0 ) (pixel)$&(207.66,92.73)\\
 &$d_1/H (mm)$&91.09\\
 &$(k_1,k_2)$&(0.15,-50.96)\\
 &$R (rad)$&(-0.02,-0.10,0.04)\\
 &$t (mm)$&(-15.16,-0.10,229.98)\\
\midrule
\multirow{2}*{The second step calibration}&$d_2  (mm)$&1.84\\
 &$d_1/f (mm)$&0.42\\
 \midrule
 \multirow{4}*{The futrher computation}&$d_1(mm)$&142.00\\
 &$d_f(mm)$&338.10\\
 &$d_3(mm)$&121.21\\
 &$H$&1.56\\
\bottomrule
\end{tabular}
\label{table2}
\end{table}

\par
\begin{figure}[t]
\centering
\includegraphics[width=3in]{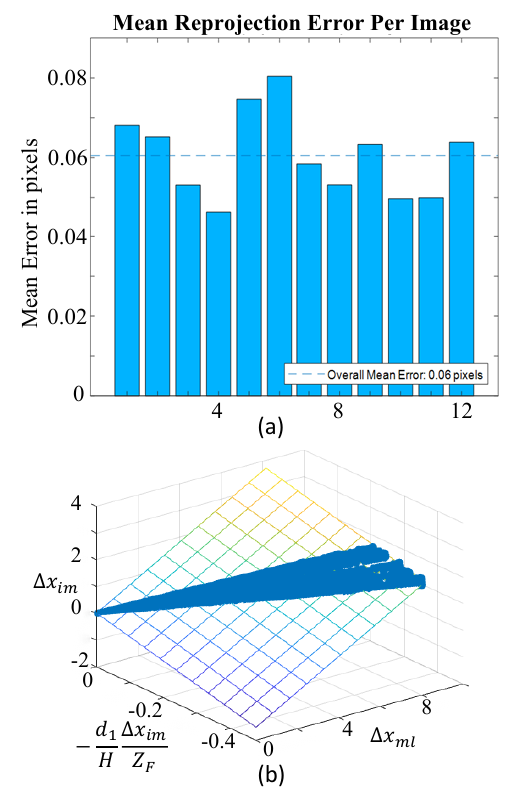}
\captionsetup{labelsep=period, font=small} 
\caption{Experimental results of the calibration. (a) Re-projection error of the corner points on the checkerboard. (b)Line fitting results.}
\label{figExpResult}
\end{figure}

To evaluate the two-step calibration performance, the re-projection errors of the corner points in the central sub-aperture images are shown in Figure \ref{figExpResult}(a), and the line fitting results are shown in Figure \ref{figExpResult}(b). As shown in Figure\ref{figExpResult}(a), the average root mean square (RMS) re-projection errors on $12$ images is about 0.06 pixels. In addition, Figure \ref{figExpResult}(b) shows the fitting relationship among $\Delta x_{ml}$, $\Delta x_{im}$, $Z_F$ of the second step calibration, and its RMS error is 0.024.

\section{Conclusion}
The light field imaging in the restrictive object space is a significant work but is easily overlooked before. Without specific considerations, the microlens image distortion will lead to not only decoding difficulty, but also wrong depth estimation (3D reconstruction).
\par
In this paper, we analyze the reason of the microlens image distortion in the ROS-LF, that is because the angular plane is changed when the traditional light field system is used in the ROS directly but without any specific considerations. To deal with the ROS-LF, we propose the flexible angular plane, the aperture constant $H$, $f$-number matching and microlens image non-distortion principles in the ROS-LF, which are all different to those in the traditional light field system. Furthermore, an advanced two-step calibration method is proposed to compute the parameters of the ROS-LF system. A simulated ROS-LF imaging system is designed to validate the principles and calibration method.
\par

\section*{Acknowledgments}
The authors thank the editors and the anonymous reviewers for their careful work and valuable suggestions for this study and declare that there is no conflict of interest regarding the publication of this paper.

\bibliographystyle{IEEEtran}
\bibliography{Draft_Light_Field_Imaging_in_the_Restrictive_Object_Space_based_on_Flexible_Angular_Plane.bib}

\begin{thebibliography}{10}
\providecommand{\url}[1]{#1}
\csname url@samestyle\endcsname
\providecommand{\newblock}{\relax}
\providecommand{\bibinfo}[2]{#2}
\providecommand{\BIBentrySTDinterwordspacing}{\spaceskip=0pt\relax}
\providecommand{\BIBentryALTinterwordstretchfactor}{4}
\providecommand{\BIBentryALTinterwordspacing}{\spaceskip=\fontdimen2\font plus
\BIBentryALTinterwordstretchfactor\fontdimen3\font minus \fontdimen4\font\relax}
\providecommand{\BIBforeignlanguage}[2]{{%
\expandafter\ifx\csname l@#1\endcsname\relax
\typeout{** WARNING: IEEEtran.bst: No hyphenation pattern has been}%
\typeout{** loaded for the language `#1'. Using the pattern for}%
\typeout{** the default language instead.}%
\else
\language=\csname l@#1\endcsname
\fi
#2}}
\providecommand{\BIBdecl}{\relax}
\BIBdecl

\bibitem{light_field_cameras}
D.~Dansereau, I.~Mahon, O.~Pizarro, and S.~Williams, ``Plenoptic flow: Closed-form visual odometry for light field cameras,'' 09 2011, pp. 4455--4462.

\bibitem{light_field_microendoscopy}
\BIBentryALTinterwordspacing
T.~M. Urner, A.~Inman, B.~Lapid, and S.~Jia, ``Three-dimensional light-field microendoscopy with a grin lens array,'' \emph{Biomed. Opt. Express}, vol.~13, no.~2, pp. 590--607, Feb 2022. [Online]. Available: \url{https://opg.optica.org/boe/abstract.cfm?URI=boe-13-2-590}
\BIBentrySTDinterwordspacing

\bibitem{Plenoptic_cameras}
\BIBentryALTinterwordspacing
F.~Dong, S.-H. Ieng, X.~Savatier, R.~Etienne-Cummings, and R.~Benosman, ``Plenoptic cameras in real-time robotics,'' \emph{The International Journal of Robotics Research}, vol.~32, no.~2, pp. 206--217, 2013. [Online]. Available: \url{https://doi.org/10.1177/0278364912469420}
\BIBentrySTDinterwordspacing

\bibitem{Convolutional_Networks}
S.~Heber and T.~Pock, ``Convolutional networks for shape from light field,'' in \emph{2016 IEEE Conference on Computer Vision and Pattern Recognition (CVPR)}, 2016, pp. 3746--3754.

\bibitem{levoy_light_1996}
\BIBentryALTinterwordspacing
M.~Levoy and P.~Hanrahan, ``Light {Field} {Rendering},'' in \emph{Proceedings of the 23rd {Annual} {Conference} on {Computer} {Graphics} and {Interactive} {Techniques}}, ser. {SIGGRAPH} '96.\hskip 1em plus 0.5em minus 0.4em\relax New York, NY, USA: Association for Computing Machinery, 1996, pp. 31--42. [Online]. Available: \url{https://doi.org/10.1145/237170.237199}
\BIBentrySTDinterwordspacing

\bibitem{Ng's_thesis}
R.~Ng, ``Digital light field photography,'' 01 2006.

\bibitem{calibration_method1}
Q.~Zhang, C.~Zhang, J.~Ling, Q.~Wang, and J.~Yu, ``A generic multi-projection-center model and calibration method for light field cameras,'' \emph{IEEE Transactions on Pattern Analysis and Machine Intelligence}, vol.~41, no.~11, pp. 2539--2552, 2019.

\bibitem{calibration_method2}
D.~Cho, M.~Lee, S.~Kim, and Y.-W. Tai, ``Modeling the calibration pipeline of the lytro camera for high quality light-field image reconstruction,'' in \emph{2013 IEEE International Conference on Computer Vision}, 2013, pp. 3280--3287.

\bibitem{calibration_method3}
\BIBentryALTinterwordspacing
H.~Duan, L.~Mei, J.~Wang, L.~Song, and N.~Liu, ``A new imaging model of lytro light field camera and its calibration,'' \emph{Neurocomputing}, vol. 328, pp. 189--194, 2019, chinese Conference on Computer Vision 2017. [Online]. Available: \url{https://www.sciencedirect.com/science/article/pii/S0925231218309615}
\BIBentrySTDinterwordspacing

\bibitem{dansereau_decoding_2013}
D.~G. Dansereau, O.~Pizarro, and S.~B. Williams, ``Decoding, {Calibration} and {Rectification} for {Lenselet}-{Based} {Plenoptic} {Cameras},'' in \emph{2013 {IEEE} {Conference} on {Computer} {Vision} and {Pattern} {Recognition}}, 2013, pp. 1027--1034.

\bibitem{zhang_generic_2018}
Q.~Zhang, C.~Zhang, J.~Ling, Q.~Wang, and J.~Yu, ``A {Generic} {Multi}-{Projection}-{Center} {Model} and {Calibration} {Method} for {Light} {Field} {Cameras},'' \emph{IEEE Transactions on Pattern Analysis and Machine Intelligence}, vol.~41, pp. 2539--2552, 2018.

\bibitem{schambach_microlens_2020}
M.~Schambach and F.~P. León, ``Microlens {Array} {Grid} {Estimation}, {Light} {Field} {Decoding}, and {Calibration},'' \emph{IEEE Transactions on Computational Imaging}, vol.~6, pp. 591--603, 2020.

\bibitem{Zhou_EPI}
\BIBentryALTinterwordspacing
P.~Zhou, Z.~Yang, W.~Cai, Y.~Yu, and G.~Zhou, ``Light field calibration and 3d shape measurement based on epipolar-space,'' \emph{Opt. Express}, vol.~27, no.~7, pp. 10\,171--10\,184, Apr 2019. [Online]. Available: \url{https://opg.optica.org/oe/abstract.cfm?URI=oe-27-7-10171}
\BIBentrySTDinterwordspacing

\bibitem{2017_EPI}
Y.~Zhang, H.~Lv, Y.~Liu, H.~Wang, X.~Wang, Q.~Huang, X.~Xiang, and Q.~Dai, ``Light-field depth estimation via epipolar plane image analysis and locally linear embedding,'' \emph{IEEE Transactions on Circuits and Systems for Video Technology}, vol.~27, no.~4, pp. 739--747, 2017.

\bibitem{shared_EPI}
G.~Wu, Y.~Liu, Q.~Dai, and T.~Chai, ``Learning sheared epi structure for light field reconstruction,'' \emph{IEEE Transactions on Image Processing}, vol.~28, no.~7, pp. 3261--3273, 2019.

\bibitem{Liu:17}
\BIBentryALTinterwordspacing
J.~Liu, D.~Claus, T.~Xu, T.~Ke{\ss}ner, A.~Herkommer, and W.~Osten, ``Light field endoscopy and its parametric description,'' \emph{Opt. Lett.}, vol.~42, no.~9, pp. 1804--1807, May 2017. [Online]. Available: \url{https://opg.optica.org/ol/abstract.cfm?URI=ol-42-9-1804}
\BIBentrySTDinterwordspacing

\bibitem{EPI_Patch}
W.~Zhou, L.~Liang, H.~Zhang, A.~Lumsdaine, and L.~Lin, ``Scale and orientation aware epi-patch learning for light field depth estimation,'' in \emph{2018 24th International Conference on Pattern Recognition (ICPR)}, 2018, pp. 2362--2367.

\bibitem{Yang_distortionless}
Y.~Shangpeng, X.~Chengcai, J.~Yifan, Z.~Guangquan, and Z.~Ping, ``Distortionless condition for microlens images in light field imaging,'' \emph{Chinese Journal of Liquid Crystals and Displays}, vol.~38, no. 829-834, 2023.

\bibitem{two_step_ZHOU2019190}
\BIBentryALTinterwordspacing
P.~Zhou, W.~Cai, Y.~Yu, Y.~Zhang, and G.~Zhou, ``A two-step calibration method of lenslet-based light field cameras,'' \emph{Optics and Lasers in Engineering}, vol. 115, pp. 190--196, 2019. [Online]. Available: \url{https://www.sciencedirect.com/science/article/pii/S0143816618309254}
\BIBentrySTDinterwordspacing

\bibitem{Zhang_calibration}
Z.~Zhengyou, ``A flexible new technique for camera calibration.'' \emph{IEEE Transactions on Pattern Analysis \& Machine Intelligence}, 2000.

\bibitem{dansereau2015linear}
D.~G. Dansereau, O.~Pizarro, and S.~B. Williams, ``Linear volumetric focus for light field cameras,'' \emph{ACM Transactions on Graphics (TOG)}, vol.~34, no.~2, Feb. 2015.

\end{thebibliography}

\newpage
\section{Biography Section}

\begin{IEEEbiography}[{\includegraphics[width=1in,height=1.25in,clip,keepaspectratio]{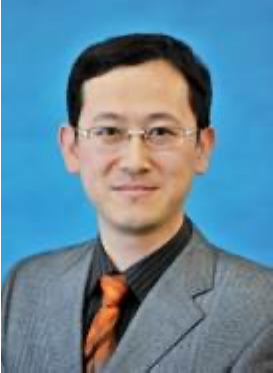}}]{Ping Zhou}
received the B.E. degree in Electronic Engineering from the University of Science and Technology of China,Hefei, China, in 2002, and the Ph.D. degree in Biomedical Engineering from University of Science and Technology of China, Hefei, China, in 2007. He is currently an associate professor with the School of Biological Science and Medical Engineering, Southeast University. His research interests include computational imaging in biomedical engineering (light field-based system and algorithm), 3D structured-light imaging in biomedical engineering, and biomedical image processing (segmentation, classification).
\end{IEEEbiography}
\begin{IEEEbiography}
[{\includegraphics[width=1in,height=1.25in,clip,keepaspectratio]{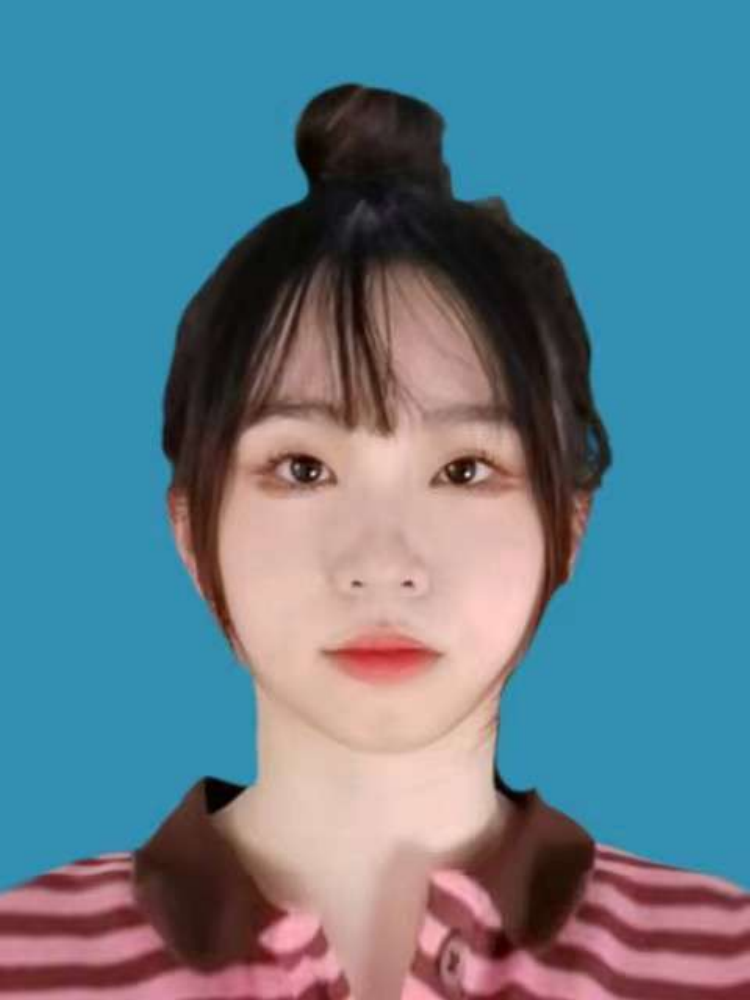}}]{Nuo Chen}
received the B.E. degree from Southeast University, Nanjing, in 2023, where she is a M.Phil student in Southeast University. Her main research interests include image processing and computer vision.
\end{IEEEbiography}
\begin{IEEEbiography}
[{\includegraphics[width=1in,clip,keepaspectratio]{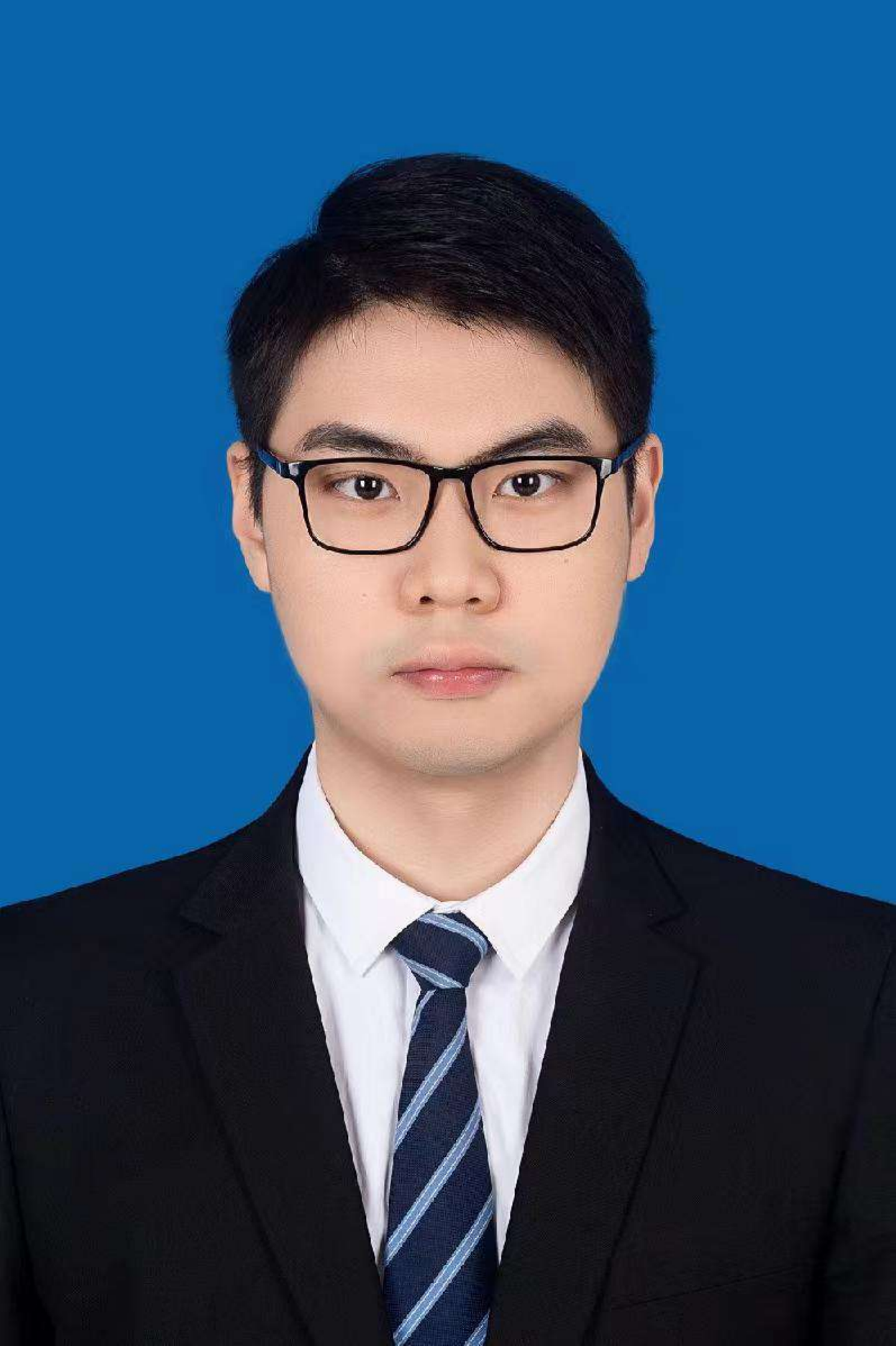}}]{Yuda Xu}
received the B.E. and M.S. degrees in Biomedical Engineering from Southeast University. His research interests include light field image processing and calibration.
\end{IEEEbiography}

\vfill

\end{document}